# Measurement of Cosmic Ray Flux in China JinPing underground Laboratory*

WU Yu-Cheng(吴昱城)[1,2]、HAO Xi-Qing(郝喜庆)[1,2;1)]、

YUE Qian(岳骞)[1,2;2)]、LI Yuan-Jing(李元景)[1,2]、

CHENG Jian-Ping(程建平)[1,2]、KANG Ke-Jun(康克军)[1,2]、

CHEN Yun-Hua(陈云华)[3]、LI Jin(李金)[1,2]、LI Jian-Min(李荐民)[1,2]、

LI Yu-Lan(李玉兰)[1,2]、LIU Shu-Kui(刘书魁)[4]、MA Hao(马豪)[1,2]、

REN Jin-Bao(任金宝)[1,2]、SHEN Man-Bin(申满斌)[3]、

WANG Ji-Min(王继敏)[3]、WU Shi-Yong(吴世勇)[3]、

XUE Tao(薛涛)[1,2]、YI Nan(易难)[1]、ZENG Xiong-Hui(曾雄辉)[3]、

ZENG Zhi(曾志)[1,2]、ZHU Zhong-Hua(朱忠华)[3]

[1]Department of Engineering Physics, Tsinghua University, Beijing 100084, China
[2]Key Laboratory of Particle and Radiation Imaging, Tsinghua University, Ministry of Education, China
[3]Yalong River Hydropower Development Company, Chengdu, 610051, China
[4]School of Physical Science and Technology, Sichuan University, Chengdu, 610041, China

Abstract: China JinPing underground Laboratory (CJPL) is the deepest underground laboratory presently running in the world. In such a deep underground laboratory, the cosmic ray flux is a very important and necessary parameter for rare event experiments. A plastic scintillator telescope system has been set up to measure the cosmic ray flux. The performance of the telescope system has been studied using the cosmic ray on the ground laboratory near CJPL. Based on the underground experimental data taken from November 2010 to December 2011 in CJPL, which has effective live time of 171 days, the cosmic ray muon flux in CJPL is measured to be $(2.0\pm0.4) \times 10^{-10}/(cm^2 \cdot s)$. The ultra-low cosmic ray background guarantees CJPL's ideal environment for dark matter experiment.

Received XX, xXX, 2012
* Supported by National Natural Science Foundation of China(10935005,11055002,11075090)
1)Email: haoxq@ihep.ac.cn
2)Email: yueq@tsinghua.edu.cn

## 1. Introduction

China JinPing underground Laboratory (CJPL) [1] locates at the middle site of a traffic tunnel under Jinping Mountain in Sichuan Province, southwest China. The length of the tunnel is about 17.5km, and the rock overburden at CJPL is about 2400m vertically, which makes CJPL the deepest underground laboratory formally running all over the world (see Fig.1) [2]. To carry out rare event experiments in CJPL, such as dark matter search, double beta decay, neutrino oscillation, et al., it is essential for us to understand the radioactive background including cosmic ray flux as well as environmental neutron and gamma ray flux. Measurements of these backgrounds will provide several important parameters for design of these possible low background experiments. This work mainly focuses on the measurement of the cosmic ray muon flux in CJPL.

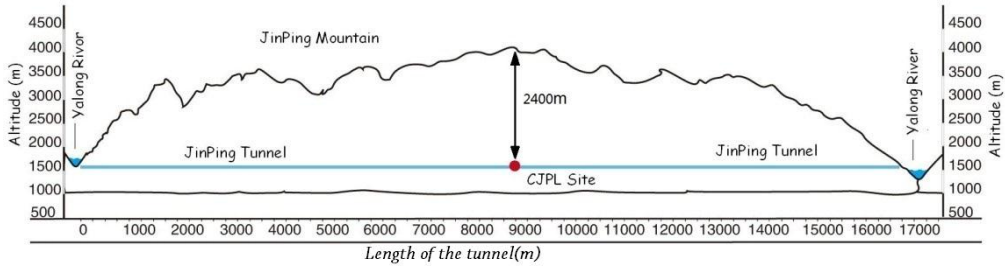

Fig. 1 The cross section of Jinping Mountain and the location of CJPL.

## 2. Detector system

In order to measure the cosmic ray muon flux in CJPL, we have built a telescope system consisting of 6 plastic scintillation detectors. Each detector has the size of 1m×0.5m×0.05m, and is read out by a photon multiplier tube (PMT, Hamamatsu CR136-01) coupled on one end. The telescope system is divided into two groups: A and B. Each group is composed of three detectors, which align vertically on a wooden shelf (see the left of Fig. 2). The distance between neighboring detectors is about 20 cm.

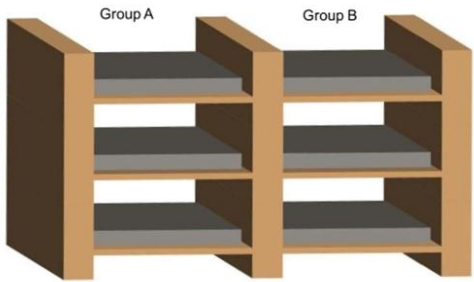 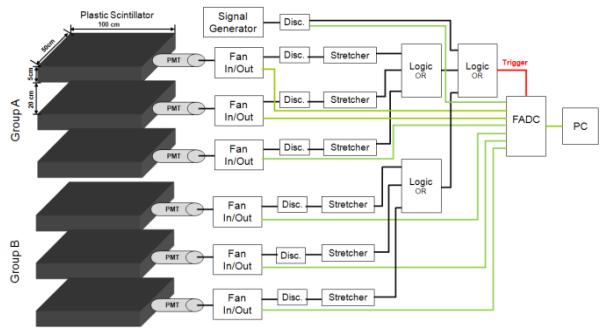

Fig. 2 Schematic layout of cosmic ray muon detector system. The left figure depicts the plastic scintillation detectors on a wooden shelf; the right figure illustrates the schematic diagram of electronics system.

The electronics system setup is illustrated in the right of Fig. 2. Signal from each PMT is firstly fanned out into two channels. One is directly fed into a FADC module (CAEN V1724, 100MHz sampling frequency) for digitization, while the other one is discriminated, stretched and then sent to logic OR modules to generate system trigger gate. Meanwhile, a signal generator, which serves as random trigger for dead time study, also contributes system trigger gate. When any detector or signal generator has signal, system trigger gate is produced. When the trigger gate arrives, the FADC module converts the whole analog pulses of all six channels to digital signals. The digital pulses are then written into hard disk for offline analysis by the DAQ program on a normal PC.

## 3. Performance of detector system

Due to the expected ultra-low cosmic ray muon flux inside CJPL, the telescope system was firstly tested using cosmic rays in a ground laboratory nearby. Pulse Shape Discrimination (PSD) method has been developed to select muon incident events, and the detection efficiencies of all detectors has also been derived. The cosmic ray muon flux of the ground laboratory will be reported as well.

In PSD method, several characteristic parameters of a pulse shape have been devised to identify the muon signal from the noises or gamma background, such as pulse amplitude, total charge (area of the pulse within 2.5μs), rise time (time rising from 10% to 90% of pulse amplitude), fall time (time falling from 90% to 10% of pulse amplitude) and so on. The first two parameters correspond to the deposited energy and their distributions represent the energy deposition spectra.

Muon particle loses its energy by ionization and radioactive processes. It will definitely deposits a large amount of energy whenever passing through plastic scintillation detector. Moreover, the angular distribution of muons at the ground is proportional to $\cos^2\theta$ [3], where θ is the zenith angle of incident muon. The muon events can be selected by two restrictions: pulse shape (pulse amplitude, rise time and fall time) constraint and triple-coincidence constraint.

According to Ref. [3], muon stopping power for Minimum Ionizing Particle (MIP) in plastic scintillator is 1.956 MeV/g/cm$^2$. So given the thickness of 5cm, the minimum energy deposition of muon penetrating from the upper surface to the lower surface of one detector is about 10 MeV, which is much higher than the ambient gamma energy(< 3 MeV). In Fig.3, distributions of pulse amplitude of muon events follow Landau distribution as theoretical prediction. Cut threshold is then set to 5000 FADC unit, right below the leading edge of Landau distribution. Events with pulse amplitude higher than 5000 FADC unit are selected for downstream analysis.

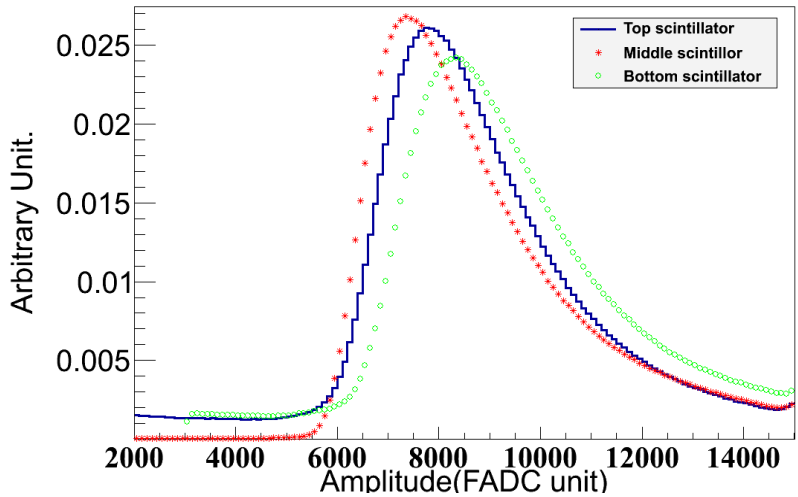

Fig. 3 Distributions of the pulse amplitude (corresponding to energy deposition spectra) of the top scintillator (blue solid line), middle scintillator (red star marker) and bottom scintillator (green circle marker) from group B. All distributions are normalized to 1 in the range of 5500~15000 FADC unit.

The rise time and fall time distributions of the triple-coincident events are plotted in Fig.4. The acceptance regions for rise time and fall time are 20-60ns and 420-600ns respectively. Noise events that fall out of these regions are thus discarded.

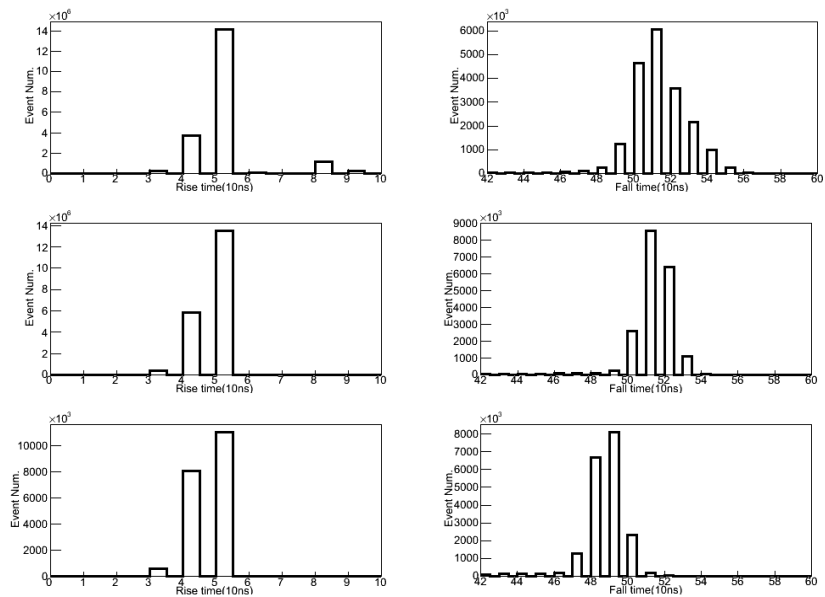

Fig. 4 Distribution of rise time (left figure) and fall time (right figure).

Restriction of triple-coincidence is defined as that pulses from all the 3 detectors in the same group should pass through the pulse shape constraint and the amplitude threshold selection. This restriction is very powerful that most of the background events can be rejected.

With all aforementioned selections, cosmic ray muon flux in the ground laboratory can be derived. Four kinds of corrections, however, are need beforehand.

The first one is dead time correction. In the right of Fig.2, the signal generator, serving as random trigger, produces periodic pulses which are independent of physics events. Dead time correction factor is then defined as the ratio of recorded number to generated number of random trigger events, which is measured to be higher than 99.9%.

The second correction is edge effect correction. In some instances, muon particles of large zenith angle would pass the edge of the top or bottom scintillators but penetrate the whole thickness of the middle one. In such cases, side detectors have less deposited energy because of the short trace of muon left in the detectors. So the muon spectra of side detectors have more events than the middle one below the leading edge of Landau distribution (see Fig. 3). Such effect is called "edge effect", and its correction factor has been calculated by Monte Carlo simulation, which is 93.4%.

The third correction is detection efficiency correction. The detection efficiency of detector $i$ is defined as:

$$\varepsilon_i = \frac{N_3}{N_2} \quad (1)$$

where $N_3$ is the number of triple-coincident events and $N_2$ is the number of double-coincident events of detectors other than detector $i$. Since the top and bottom detectors both have edge effect (see Fig. 3), Eq. (1) is only valid for the detectors in the middle position. All detectors took turns in the middle position to measure its detection efficiency. The results of the 6 detectors are listed in Table 1, which are all larger than 99%.

The last correction is solid angle correction, which considers muons that have too large zenith angle to create triple-coincidence. This correction has also been studied by Monte Carlo simulation with $\cos^2$ distribution assumption of incident muons. The solid angle correction factor (of group A or group B) is defined as the proportion of the triple-coincident cosmic ray muon events to the cosmic ray muon events which pass through the top scintillator. Fig. 5 shows solid angle correction factor as a function of θ. Because the plastic scintillation detectors cannot distinguish the direction of the incident muon, so the correction factor takes the weighted average value of 33.9%, which is the ratio of the total survived triple-coincident events number to the total events number passing through the top detector.

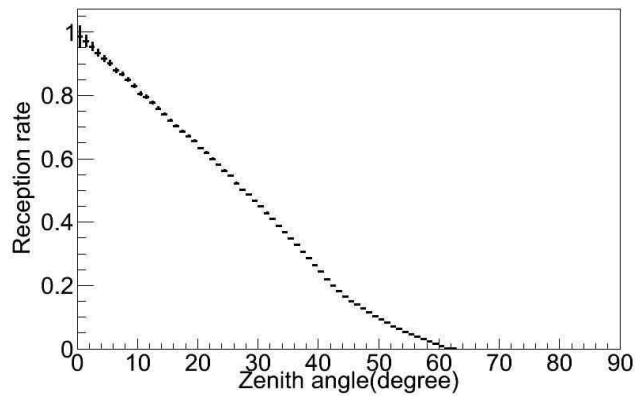

Fig. 5 Solid angle correction for telescope system. The integral value of 33.9% is used.

All the restrains and the correction results are listed in Table 1. The cosmic ray muon flux of the ground laboratory can be calculated by

$$\frac{N_3}{\varepsilon_1 \times \varepsilon_2 \times \varepsilon_3 \times \varepsilon_4 \times t} \quad (2)$$

where $N_3$ is the number of triple-coincident events, $\varepsilon_1$ - $\varepsilon_4$ are the correction factors and $t$ is the live time.

Based on more than 19 million selected cosmic ray muon events, the muon flux of the ground laboratory is measured to be 144.7 /(m² s), where the altitude is about 1600m. The result is consistent with our previous measurement [4] and others' result [3] at the same altitude.

## 4. Measurement of Cosmic ray muon flux in CJPL

After testing in the ground laboratory, the telescope system was moved and set up in CJPL. The detectors'

position as well as the electronics setup was kept the same as ground measurement.

Table 1 Restrains and correction factors of the detector system.

| | |
|---|---|
| Rise time restrain | 20-60 ns |
| Fall time restrain | 420-600 ns |
| Amplitude threshold restrain | 5000 FADC unit |
| Dead time correction factor | > 99.99% |
| Edge effect correction factor | 93.4% |
| Detector efficiency correction factor | 99.2%(A top), 99.3%(A middle), 99.3%(A bottom) |
| | 99.4%(B top), 99.7%(B middle), 99.0%(B bottom) |
| Solid angle correction factor | 33.9% |

Before data analysis, data quality has been verified. Runs of dataset less than 2 days are firstly removed. The stability of the pulse pedestal has also been investigated. If there are one or more abnormal jumps in the averaged pedestals, the total run is also eliminated. From November 2010 to December 2011, we have accumulated 231 days' data. In the whole 54 runs, there remain 14 runs after quality check, corresponding to live time of 171 days.

Muon events are selected by the same selection criteria as ground measurement. After pulse shape and triple-coincidence selection, we get 28 events from two groups. To get better statistics, pulses of selected muon events from all 6 detectors of the two groups are counted up to obtain a general amplitude distribution. In Fig. 6 the general amplitude distributions of both ground and underground measurements are drawn together. The small bump around 15500 FADC units is caused by the saturation of FADC module. It is due to the outrange signal stack of high energy cosmic ray or superposition of coincident high energy radiations. Although limited by statistical errors, the underground spectrum is in line with the ground spectrum. One can see that there is no obvious difference between them.

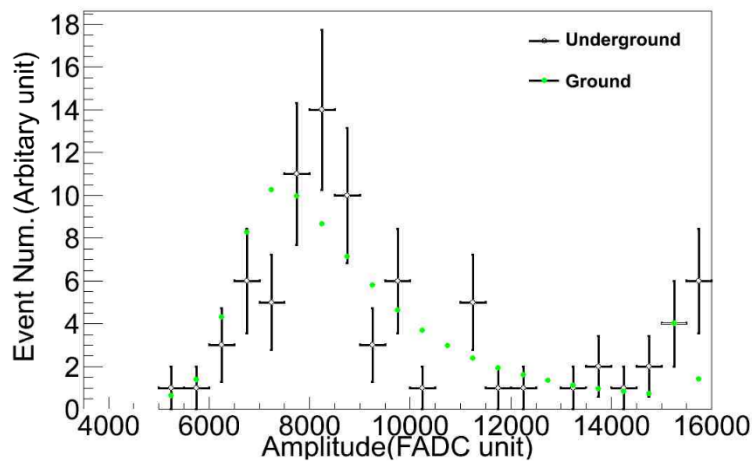

FIG 6 Pulse amplitude distributions of underground (black dot with error bar) and ground (green dot) measurements. For better statistics, selected muon pulses from all 6 detectors are taken into account.

Corrections are still needed to calculate the cosmic ray muon flux. Dead time correction has been measured by random trigger as before. Detection efficiencies of all detectors are chosen from the ground measurement. From Fig. 1 we can see that muons penetrating into CJPL have the shortest paths in the vertical direction, and muons in other directions have longer paths that are more likely to decay or to be stopped by rocks. So the angular distribution of cosmic ray muons in CJPL is much more vertical-tended than ground situation. Because we do not know the angle distribution, its correction factor is treated as 1 for this underground measurement and the edge effect is ignored.

After all the corrections, the cosmic ray muon flux in CJPL is measured to be $(2.0\pm0.4)\times10^{-10}$ /(cm$^2$ s), or $61.7 \pm 11.7$ /(m$^2$ year). Considering the large statistic uncertainty of 18.9%, we neglect the systematic error.

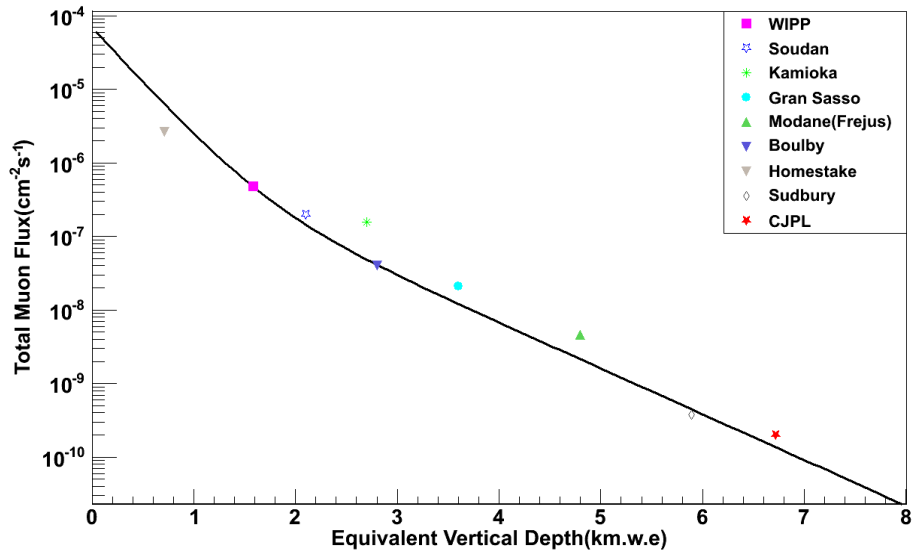

Fig. 7 The flat-earth model [5] describing muon flux v.s. depth of the main underground laboratories [6-13]. For CJPL, the water equivalent depth is computed with the rock density 2.8g/cm$^3$ and depth 2400m.

Fig. 7 shows our result comparing with flat-earth model [5] which describes the cosmic ray muon flux of main deep underground laboratories (i.e., WIPP [6], Soudan [7], Kamioka [8], Gran Sasso [9], Modane (Frejus) [10], Boulby [11], Homestake [12], Sudbury [13]) at different depths. The water equivalent depth of CJPL is calculated to be 6720m as the product of the measured average rock density of 2.8g/cm$^3$ and the depth of 2400m. This result of CJPL is consistent with the flat-earth model.

## 5. Summary

A telescope system of 6 plastic scintillation detectors and electronics setup has been set up to measure the cosmic ray muon flux in CJPL. The system has been tested in the ground laboratory nearby, and its performances have been studied. Data analysis method has also been developed for event selection and flux correction. With 171 effective data, the cosmic ray muon flux in CJPL is measured to be $(2.0\pm0.4)\times10^{-10}$ /(cm$^2$ s) at the depth of 6720 m.w.e (water equivalent meter), which is in line with the flat-earth model. The ultra-low background provides the necessary condition of environment for rare event experiments, such as the operating dark matter direct detection.